\documentclass[twocolumn]{revtex4-2}

\usepackage[T1]{fontenc}
\usepackage{amsmath,amssymb,amsfonts}
\usepackage{algorithmic}
\usepackage{graphicx}
\usepackage{textcomp}
\usepackage{xcolor}
\def\BibTeX{{\rm B\kern-.05em{\sc i\kern-.025em b}\kern-.08em
    T\kern-.1667em\lower.7ex\hbox{E}\kern-.125emX}}

\begin{document}

\title{Hybrid Classical-Quantum Supercomputing: A demonstration of a multi-user, multi-QPU and multi-GPU environment}

\author{Mateusz Slysz$^1$}
\author{Piotr Rydlichowski$^1$}
\author{Krzysztof Kurowski$^1$}
\author{Omar Bacarreza$^3$}
\author{Esperanza Cuenca Gomez$^2$}
\author{Zohim Chandani$^2$}
\author{Bettina Heim$^2$}
\author{Pradnya Khalate$^2$}
\author{William R. Clements$^3$}
\author{James Fletcher$^3$}
\email{jfletcher@orcacomputing.com}
\affiliation{$^1$Poznanskie Centrum Superkomputerowo-Sieciowe \\
$^2$NVIDIA\\
$^3$ORCA Computing}

\begin{abstract}
Achieving a practical quantum advantage for near-term applications is widely expected to rely on hybrid classical-quantum algorithms. To deliver this practical advantage to users, high performance computing (HPC) centers need to provide a suitable software and hardware stack that supports algorithms of this type. In this paper, we describe the world's first implementation of a classical-quantum environment in an HPC center that allows multiple users to execute hybrid algorithms on multiple quantum processing units (QPUs) and GPUs. Our setup at the Poznan Supercomputing and Networking Center (PCSS) aligns with current HPC norms: the computing hardware including QPUs is installed in an active data center room with standard facilities; there are no special considerations for networking, power, and cooling; we use Slurm for workload management as well as the NVIDIA CUDA-Q extension API for classical-quantum interactions. We demonstrate applications of this environment for hybrid classical-quantum machine learning and optimisation. The aim of this work is to provide the community with an experimental example for further research and development on how quantum computing can practically enhance and extend HPC capabilities.
\end{abstract}

\maketitle

\section{Introduction}

Hybrid classical-quantum ("hybrid") algorithms are widely expected to show the earliest practical advantage for quantum computers in non-fault-tolerant architectures \cite{mcclean2016theory,endo2021hybrid,callison2022hybrid,beck2024integrating,elsharkawy2024integration}. Theoretical modeling and scenario-based analyses provide strong evidence for benefits from integrating HPC and quantum computing, while highlighting the scarcity of experimental testbeds and empirical datasets \cite{honda2025advantages}. Looking to the future, fault-tolerant quantum computers will also need classical compute for many reasons including hardware control processes, pre- and post-processing of data, providing the user interface, scheduling jobs and monitoring health and performance. 

However, experimental realizations of a realistic quantum-integrated HPC cluster have been lacking. Architectures such as \cite{shehata2025software} and software elements such as \cite{mantha2024pilot} have been proposed to meet the need for classical-quantum integration, and developers have delivered software implementations of hybrid algorithms such that research can now be done on the body of software itself \cite{zappin2024quantum,bensoussan2025taxonomy}. Despite this progress, there is still a gap between the current state of the art and a demonstration of a realistic cluster, which we define as one that allows us to demonstrate and explore system-level behavior incorporating multiple users, scheduling, standard software libraries and network interfaces across multiple CPUs, GPUs and QPUs. 

In this paper, we describe our implementation of a quantum-integrated HPC cluster at the Poznan Supercomputing and Networking Center (PCSS). Our approach of building a realistic cluster by extending an existing active HPC cluster has numerous benefits. These include: providing a testbed for co-design of classical-quantum algorithms; proving out a heterogeneous architecture; enabling collaborative processes across organizations; a baseline CUDA-Q photonics implementation for future evolution; developing user community confidence to work in this paradigm; practical experiments in optimizing across hardware and software for hybrid algorithms; and a platform starting to provide abstraction layer for end users.

We also demonstrate several hybrid algorithms running on this HPC environment. These algorithms provide examples of variational algorithms \cite{mcclean2016variational} characterized by iterations over a classical loop that calls a quantum subroutine. Although further research is required to identify routes to quantum advantage \cite{gujju2024quantum}, the success of these initial applications suggests that quantum machine learning and optimization with photonic quantum processors can be a practical proposition. These demonstrations serves to illustrate the type of workflow supported by this HPC environment, and shows that this environment can be used to solve real-world problems. We finally suggest a range of further avenues to explore using this setup.

\section{Background}

HPC environments are complex systems composed of shared CPU, GPU and, incipiently, QPU-based resources that must be efficiently allocated to multiple end-users simultaneously. In such environments, many different users run often highly variable workloads, ranging from short tasks to long-running simulations, and may require support for Message Passing Interface (MPI) libraries, GPU acceleration, and strict dependency management. Effective job scheduling is therefore central to HPC operations, with fair-share scheduling emerging as a critical strategy to balance equitable access and optimal resource utilization among competing users and groups. Schedulers such as Slurm \cite{yoo2003slurm}, Flux, LSF, and PBS Pro enable this by offering advanced policies and algorithms that administrators can use to define priorities, enforce usage quotas, and manage conflicting demands based on historical usage and other criteria \cite{Kurowski2004}. Understanding user workload patterns is essential, as requests may span single-core jobs to highly parallel applications utilizing numerous CPUs or GPUs. Although there are established best practices for scheduling CPU and GPU jobs, the integration of quantum processing units (QPUs) into HPC infrastructures introduces new challenges \cite{britt2017high}. These include adapting the shared resource model to account for the fundamentally different characteristics of quantum workloads, which necessitates urgent development of novel resource scheduling and job management strategies, as well as efficient data movement between classical and quantum resources to maintain high resource utilization and user satisfaction in next-generation QC-HPC environments.

\subsection{HPC at PCSS}

The new hybrid multi-QPU testbed, established as part of a large-scale HPC setup, is surrounded by dedicated NVIDIA GPU-accelerated nodes available within the ALTAIR and PROXIMA systems at PCSS. The ALTAIR system comprises nine GPU nodes, each configured with eight NVIDIA V100 GPUs, offering 32 GB of memory per GPU. This architecture is well-suited for parallel computing workloads commonly encountered in artificial intelligence (AI) applications. In contrast, the PROXIMA system offers substantially greater computational capacity, consisting of eighty-seven GPU nodes, each equipped with four NVIDIA H100-94 SXM5 GPUs, each providing 94 GB of HBM2e memory. The advanced architecture and significantly larger memory footprint of the H100 GPUs in PROXIMA enable superior performance for highly demanding AI workloads, including large-scale model training and high-throughput inference workloads generated by different users and groups.

\subsection{Photonic quantum processors}

\begin{figure}[ht]
    \centering
    \includegraphics[width=1\linewidth]{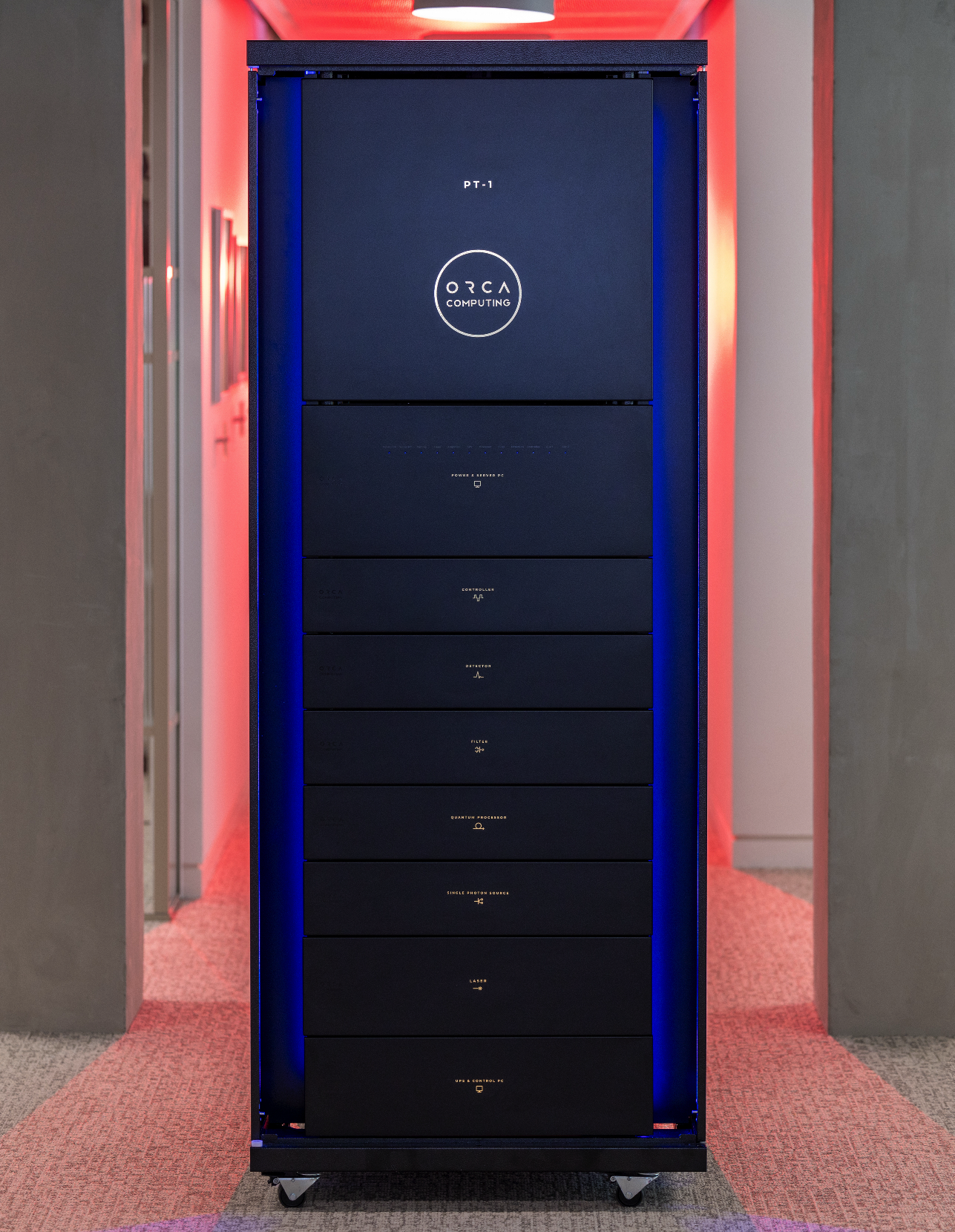}
    \caption{An ORCA Computing PT-1 system. The PT-1 is a first-generation quantum system operating at room temperature and supporting quantum optical circuits with up to 4 photons in 8 qumodes.}
    \label{fig:pt1}
\end{figure}

The quantum processors used for this environment are two ORCA Computing PT-1 systems installed at PCSS in December 2023. The PT-1 is a room-temperature photonic quantum processor with a 19" single-cabinet form factor as shown in figure \ref{fig:pt1}. It consists of a heralded single photon source, an interference circuit implemented by two sequential optical fiber delay lines with programmable coupling parameters \cite{novak2024boundaries}, and single photon avalanche photodiodes that are multiplexed to provide photon number resolution. Its average power consumption is roughly 600W. As a first-generation quantum system, the PT-1 can support up to 4 photons interfering in 8 optical modes or "qumodes". One of the two PT-1 systems was installed in the PCSS datacentre room, and the second system was installed in an optics laboratory in the same building. Connections to the PCSS network are provided via Ethernet.

A specificity of near-term photonic quantum processors, including the PT-1, is that unlike many other modalities quantum circuits are not constructed from qubits and gates. In these systems, identical single photons are created, interfered with each other in a programmable interference circuit, and measurements are performed to determine where the photons leave the circuit \cite{aaronson2011computational}. Though this process is not universal for quantum computing, photonic platforms have some advantages compared to other modalities from the point of view of HPC integration. Physical integration can be easier thanks to room-temperature operation and the use of photonic components that are designed to operate within a datacentre environment. Photonic quantum processors have also demonstrated a unique ability to scale to real-world data in hybrid classical-quantum machine learning algorithms \cite{bacarreza2025quantum,cimini2025large}, thanks to the absence of decoherence and the support for long-range entanglement with shallow non-nearest neighbor circuits \cite{novak2024boundaries}.

The PT-1 systems already include some basic job management features. A typical job on the PT-1 involves running a circuit with a specified input state and circuit parameters for a certain number of shots. Jobs are processed sequentially via an internal FIFO queue. Jobs are submitted and results are retrieved using a set of REST APIs that are typically called by higher-level software such as the ORCA proprietary software development kit, or the NVIDIA CUDA-Q extension API. Additional admin-level APIs are also provided to monitor system status. In this work, we went beyond these existing basic functionalities and designed and developed the tools required to fully integrate these systems into an active HPC cluster.

\subsection{CUDA-Q for accelerated quantum supercomputing}

\begin{figure}[ht]
    \centering
    \includegraphics[width=1\linewidth]{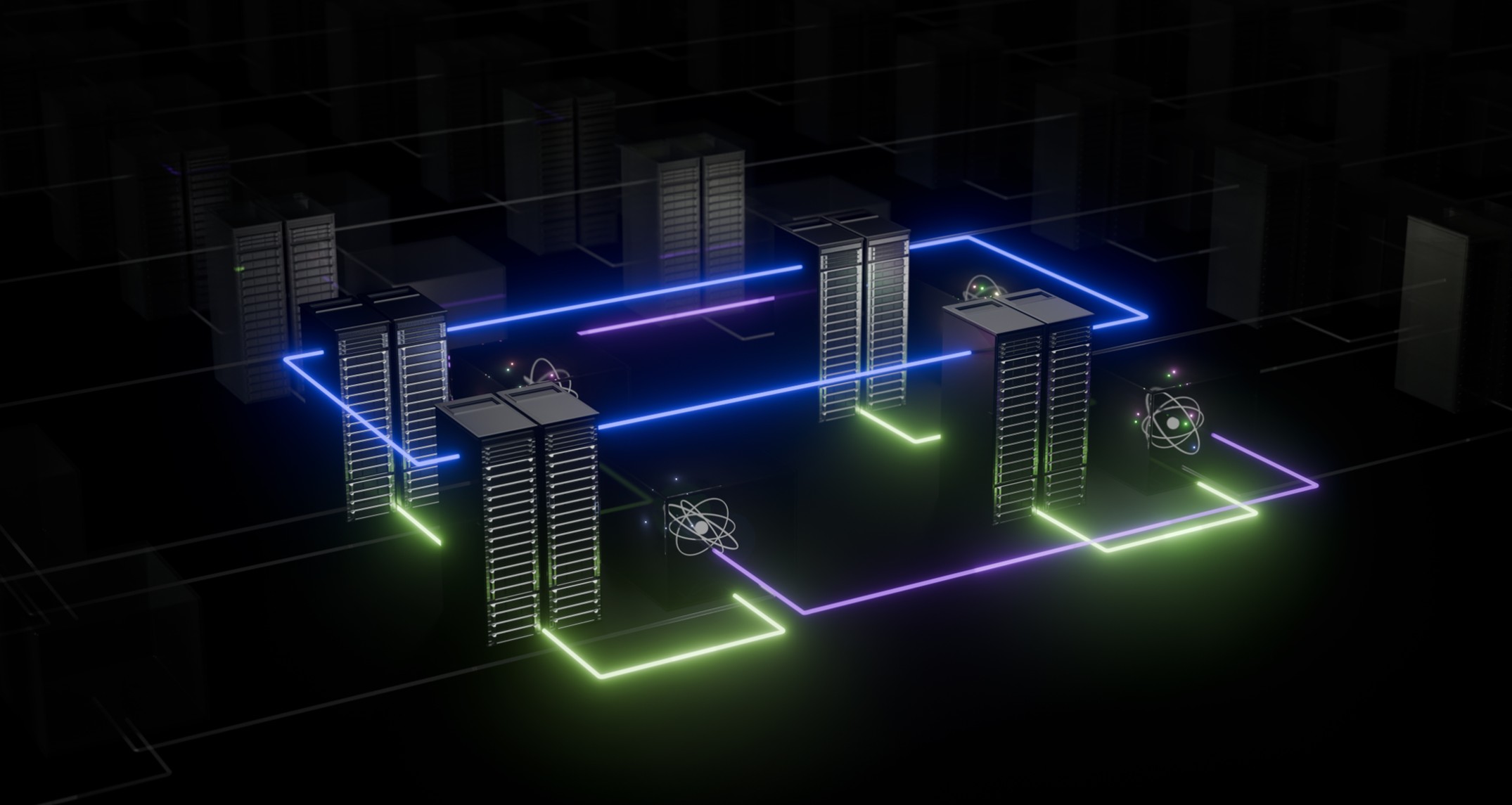}
    \includegraphics[width=1\linewidth]{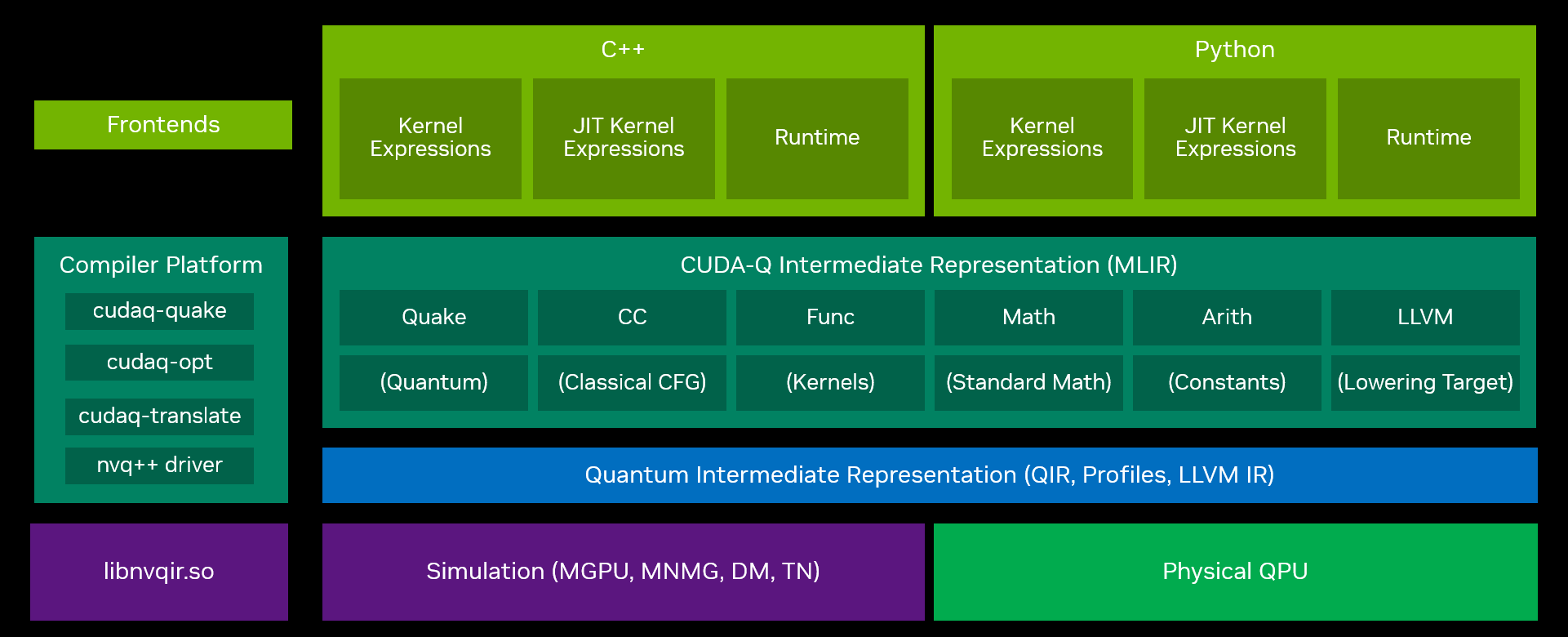}
    \caption{Top: High-level representation of an Accelerated Quantum Supercomputer. Bottom: The CUDA-Q Stack. ©NVIDIA}
    \label{fig:cudaq}
\end{figure}

In the type of heterogeneous architecture that we consider, the classical computer acts as the control plane, managing the workflow, preparing quantum states, measuring results, and performing classical post-processing. The QPU serves as a specialized accelerator for certain parts of the computation. Software integration is crucial because many promising quantum algorithms, such particularly those for optimization or quantum chemistry, are inherently hybrid, requiring iterative feedback loops between classical optimization routines and quantum circuit execution. For instance, variational quantum eigensolvers (VQE) or quantum approximate optimization algorithms (QAOA) rely heavily on classical optimizers to adjust quantum circuit parameters based on measurement outcomes.

NVIDIA CUDA-Q is a software development platform designed for these heterogeneous classical-quantum architectures, which are also referred to as "accelerated quantum supercomputers". CUDA-Q provides a unified environment for developing, simulating, and deploying hybrid algorithms. Building upon familiar CUDA programming model concepts such as kernels, and extending those to the quantum realm, CUDA-Q is an open-source quantum development platform that streamlines hybrid development with a unified programming model, is QPU agnostic and is efficiently integrated with all qubit modalities. CUDA-Q includes best-in-class compiler and runtime tools that optimize user code with backend-specific tailored performance. CUDA-Q offers interoperability with AI and HPC workflows and integration with quantum tools across the stack and enables the design and simulation of quantum systems. This interoperability with AI workflows is especially useful to accelerate high-performance AI workflows that enable hybrid applications. An overview of the CUDA-Q stack is presented in figure \ref{fig:cudaq}.

The ability of CUDA-Q to support computation on GPU, CPU, and QPU resources from within a single program, and its ability to offer GPU-accelerated simulations and high-performance AI workflows for quantum computing, makes it well suited for our hybrid HPC environment. Our implementation of this environment included developing an integration of ORCA Computing's PT-1 systems within CUDA-Q.

\section{Software and workload management}

\subsection{Photonic integration within CUDA-Q}

\begin{figure}[htp]
    \centering
    \includegraphics[width=\linewidth]{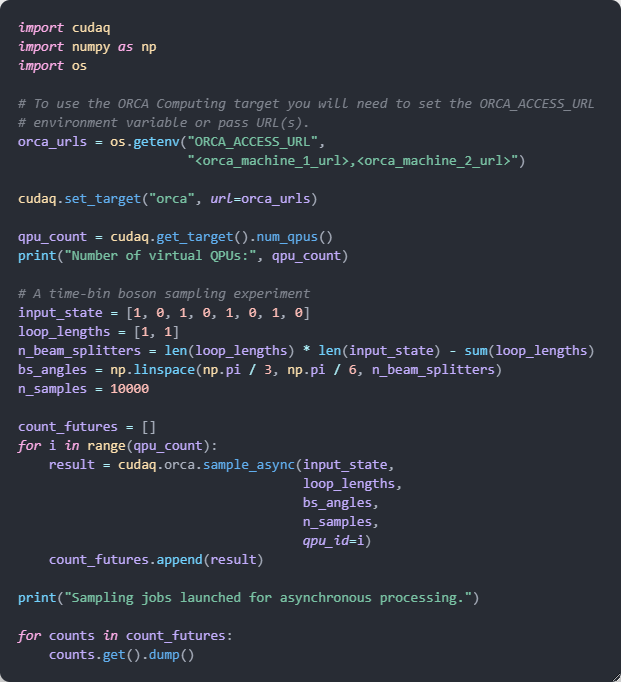}
    \includegraphics[width=\linewidth]{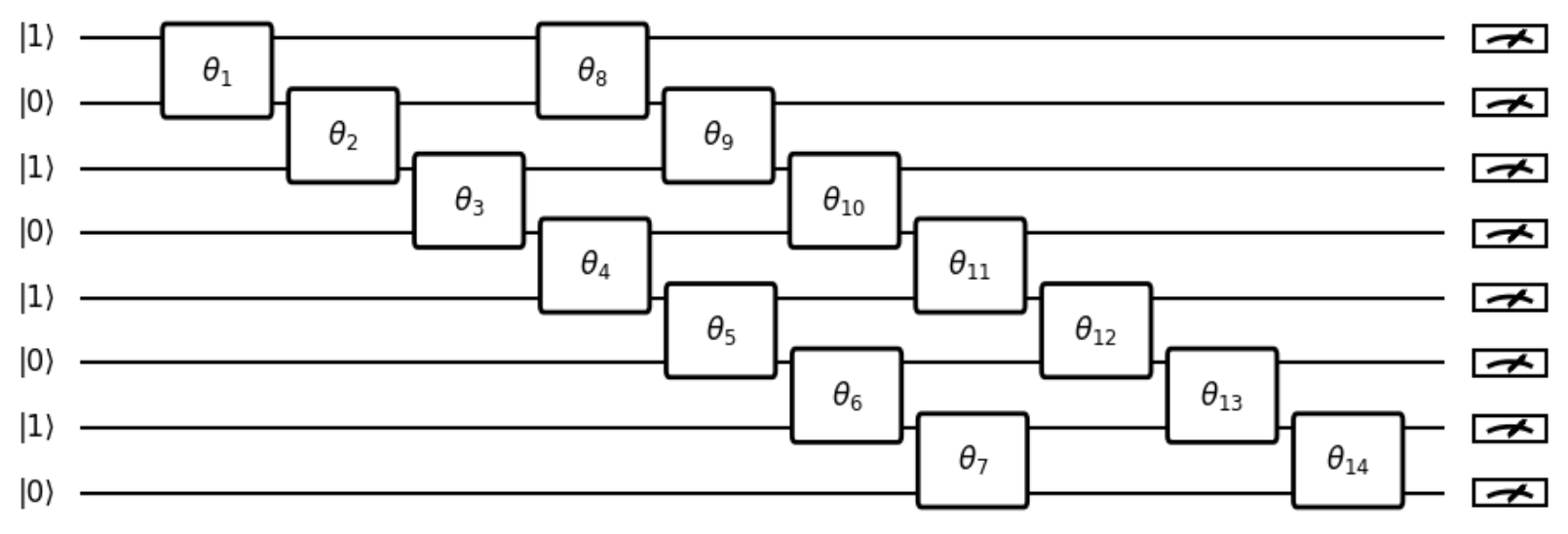}
    \caption{Top: Example code for running the simple circuit shown at bottom using two ORCA Computing PT-1 systems. The circuit shown here has 4 input photons in 8 qumodes and 14 programmable parameters, which are the beam splitter coupling ratios between two qumodes.}
    \label{fig:orcascript}
\end{figure}

As part of our implementation of a hybrid HPC environment, we integrated the ORCA Computing PT-1 systems within CUDA-Q as a backend. This backend makes use of multi-QPU functionality encapsulating ORCA's photonic QPUs as independent HTTP REST server instances. CUDA-Q programmers can access the ORCA API from either C++ or Python. In addition to support for real hardware, CUDA-Q also allows users to simulate photonic quantum processors.

Figure \ref{fig:orcascript} shows an example of how to use CUDA-Q to run a simple circuit on two PT-1 systems. In this example, 4-photon input states defined in \verb|input_state| are created and sent into an optical circuit that is physically implemented by two sequential optical delay lines of equal length defined in \verb|loop_lengths|. Each delay line adds one diagonal line of programmable beam splitters, as described in \cite{novak2024boundaries}. The programmable coupling parameters are defined in \verb|bs_angles|, and the circuit is run \verb|n_samples=10000| times. Asynchronous logic allows users to use multiple QPUs to accelerate workflows.

On top of CUDA-Q, we also developed an internal higher-level Python software stack to provide an algorithm-level programming interface. This allows users to directly implement algorithms appropriate for PT Series processors such as optimization \cite{pcss2025hybriddemo}, hybrid quantum/classical GANs \cite{bacarreza2025quantum}, or photonic quantum neural networks \cite{gan2022fock}, which we call "PTLayer" in our software implementation.

\subsection{Workload management with Slurm}

\begin{figure}[htp]
    \centering
    \includegraphics[width=\linewidth]{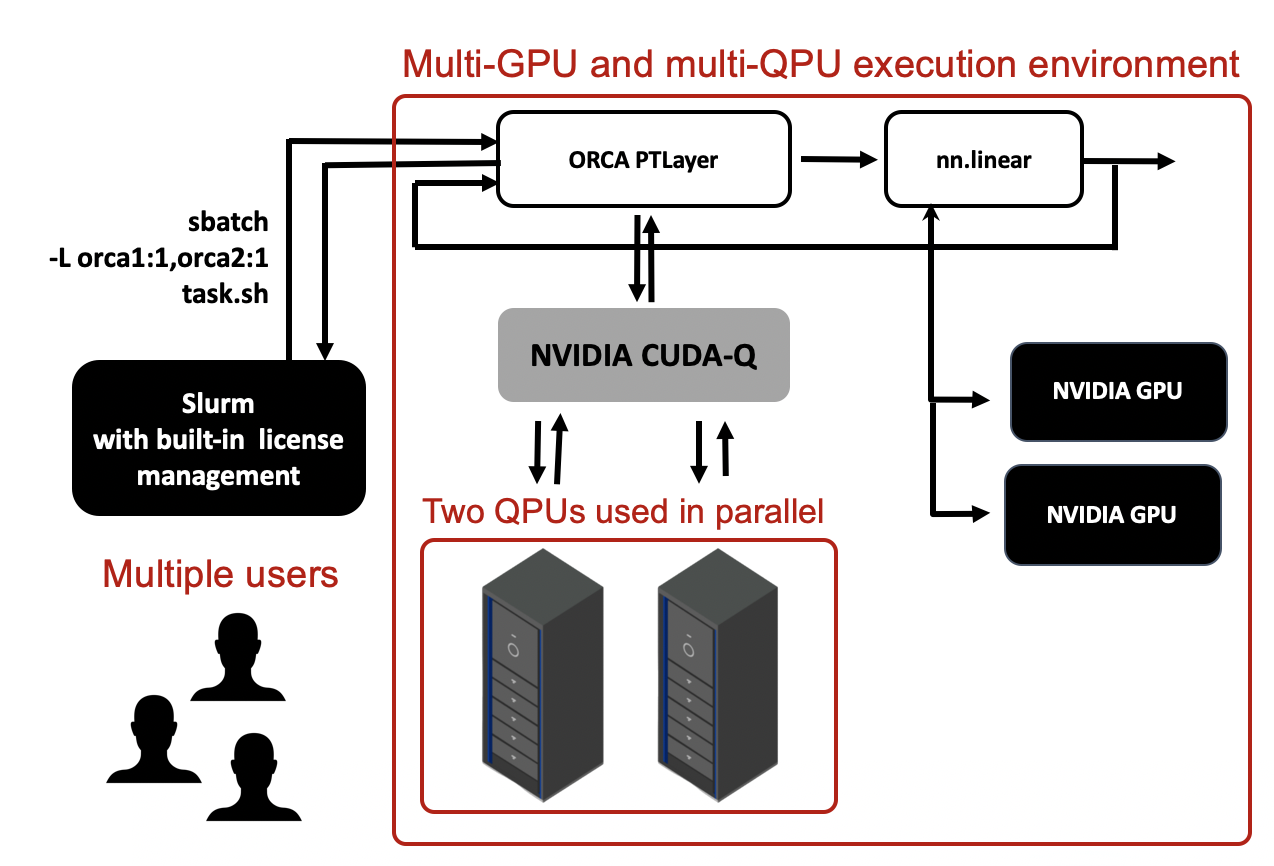}
    \caption{The overall architecture of multi-user, multi-GPU and multi-QPU integration using Slurm and NVIDIA CUDA-Q. We use a hybrid quantum neural network as an example workflow, with a quantum neural network running on the PT-1 (the "PTLayer") and a classical neural network ("nn.linear") running on GPUs.}
    \label{fig:architecture}
\end{figure}

We adopt Slurm for workload management. Many HPC environments utilize Slurm as a resource manager to coordinate computational workloads. Slurm operates on the service nodes and the compute nodes equipped with CPUs and GPUs where large-scale simulations are executed. 

We extended Slurm to enable the simultaneous allocation of heterogeneous resources, including CPUs, GPUs, and Quantum Processing Units (QPUs), which is essential for executing hybrid classical-quantum experiments. This builds on Slurm's basic license management mechanism, which is typically implemented through license counting. Once the resources have been allocated by Slurm, CUDA-Q facilitates communication between these components. This framework, illustrated in figure 4, enables GPU-based workloads running on HPC nodes to interact at runtime with remote QPUs via NVIDIA CUDA-Q, effectively integrating quantum computing into classical HPC workflows.

\section{Hybrid classical-quantum applications}

\begin{figure}[htp]
    \centering
    \includegraphics[width=0.8 \linewidth]{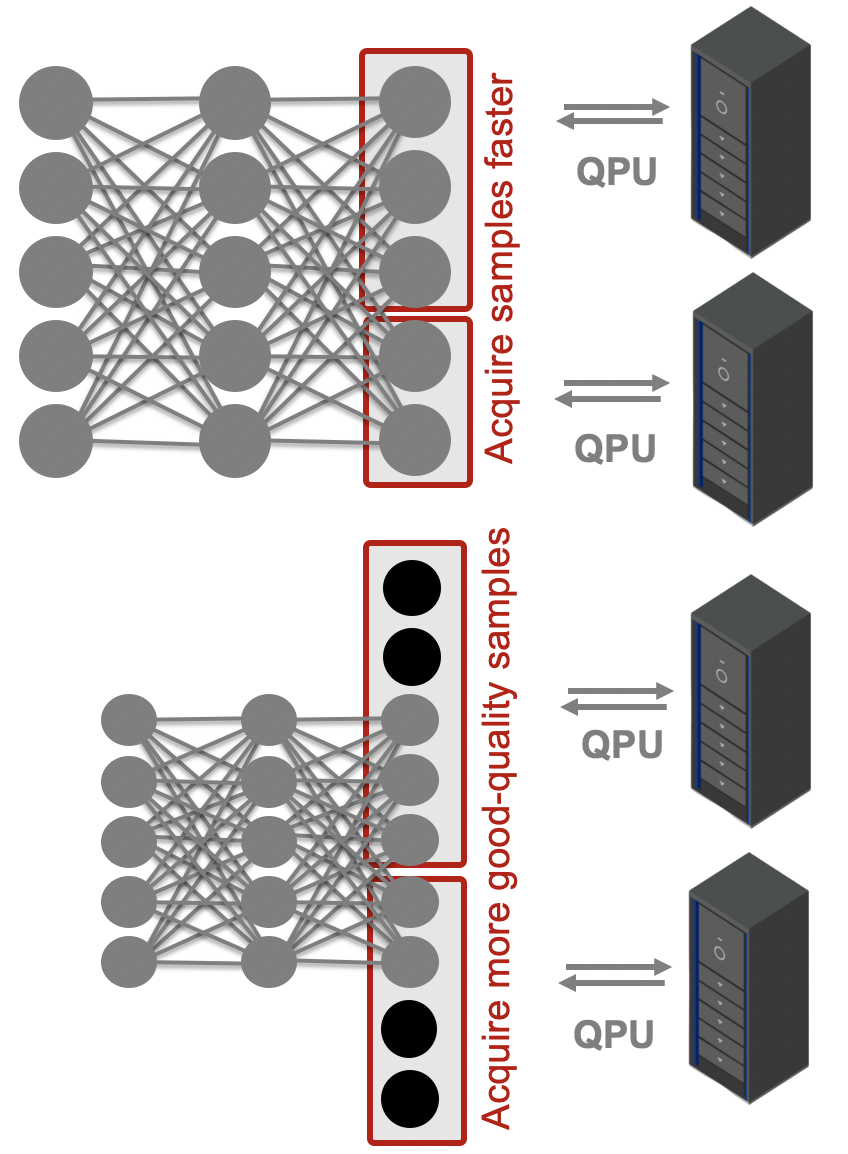}
    \caption{Possible hybrid scenarios and added value of concurrent access to multiple QPUs to accelerate hybrid machine learning algorithms. Dividing a task between two identical quantum processors can reduce wall clock time. Alternatively, for a given runtime, using both quantum processors yields twice as many samples leading to a more accurate estimate of the observable.}
    \label{fig:hybrid}
\end{figure}

Here, we discuss the advantages of running hybrid algorithms in a multi-QPU and multi-GPU environment and present some examples of such algorithms. Having access to two QPUs can accelerate hybrid quantum/classical algorithms in two ways, illustrated in figure 5 in the case of quantum neural networks. Since variational algorithms typically require estimating the expectation value of an observable over a potentially large number of samples, dividing the task between two identical quantum processors can reduce wall clock time. Alternatively, for a given runtime, using both quantum processors yields twice as many samples leading to a more accurate estimate of the observable.

We have demonstrated both optimization and quantum neural network algorithms for this environment \cite{pcss2025hybriddemo}. In optimization, we ran a Binary Bosonic Solver (BBS) algorithm on problems such as Max-Cut and the Job Shop Scheduling Problem. The BBS algorithm uses samples from photonic quantum processors as candidate solutions, which progressively improved solutions found via a classical feedback loop. Thanks to efficient encoding strategies and a "tiling" technique that divides problem instances into smaller subproblems processed in parallel, we were able to solve problem instances with up to 30 variables \cite{pcss2025hybriddemo}. In machine learning, we embedded the photonic quantum processor as a variational layer within a classical neural network. We tested this hybrid architecture on classification and representation learning tasks. While the hybrid models did not consistently surpass the accuracy of a purely classical baseline, they exhibited enhanced stability during training. These experiments illustrate how the hybrid photonic platform supports tightly coupled learning scenarios.

Another promising direction is the use of quantum processors to optimize neural network parameters. For example, we have demonstrated a "quantum neural architecture search" (QNAS) algorithm in our hybrid environment. This approach involves mapping QPU parameters to a candidate neural network architecture running on GPUs. By training the QPU parameters, we can evolve a population of candidate network architectures over successive generations, with each individual’s fitness determined by its classification accuracy on a downstream task. We performed a demonstration of such an algorithm on two datasets: the cybersecurity UNSW-NB 15 dataset \cite{cyber_benchmark} and the well-known Iris dataset \cite{pcss2025hybriddemo}. Our work used an ensemble of 10 neural network architectures, with each QPU evolving 5 architectures. As illustrated in figure 6, we found that the evolved architectures outperformed the initial and default configurations in terms of accuracy. These results point towards the potential of combining evolutionary computation with distributed classical-quantum systems to autonomously discover high-performing neural architectures in a parallel and resource-efficient manner. 

\begin{figure}[htp]
    \centering
    \includegraphics[width=\linewidth]{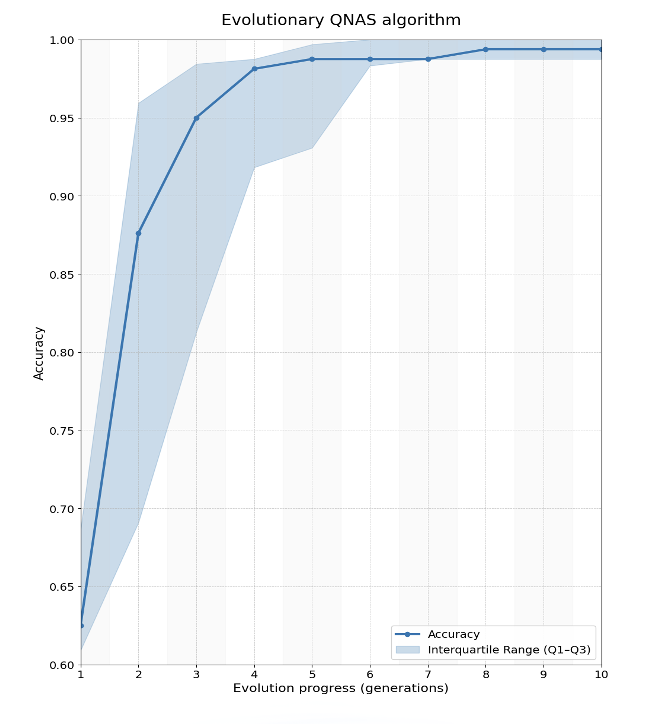}
    \caption{Experimental results obtained for a quantum neural architecture search evolutionary algorithm for optimizing neural network architectures allocated on GPUs and QPUs to solve a classification problem using the cybersecurity UNSW-NB 15 dataset. The shaded area corresponds to the standard deviation over 10 runs.}
    \label{fig:orcaexp}
\end{figure}

In addition to these demonstrations, we also highlight several other hybrid algorithms that are suitable for this experimental platform. Recent work on large-scale reservoir computing for image classification \cite{kornjavca2024large,cimini2025large,gong2025enhanced} and support vector machines \cite{yin2024experimental} highlights the advantages of a photonic platform for quantum machine learning. Hybrid generative models using photonic quantum processors combined with large-scale neural networks have been shown to scale to real-world datasets \cite{bacarreza2025quantum} and to support anomaly detection workflows \cite{kailasanathan2025quantum}. Gradient-based training methods such as \cite{facelli2024exact, hoch2025variational} that can support these applications can also be accelerated with a multi-QPU environment. In terms of quantum algorithms for optimizing the parameters of neural networks running on GPU, a notable example designed for multi-QPU environments can be found in \cite{chen2025distributed}.

These examples underscore the potential of hybrid applications in machine learning that can leverage a multi-QPU and multi-GPU environment. We anticipate that the availability of such an environment to algorithm developers will further encourage the development of new applications.

\section{Conclusion}

We have presented the first experimental deployment of a hybrid classical–quantum computing environment within an operational HPC datacenter, integrating multiple QPUs and GPUs into a multi-user, scheduler-managed cluster. Our implementation demonstrates that such a platform can be realized without specialized facility requirements, and can leverage established HPC practices, such as Slurm-based workload management, while supporting emerging hybrid algorithm workflows.

The three core components of our implementation of a practical hybrid HPC–QC platforms are: a programming framework capable of expressing and orchestrating hybrid workloads (CUDA-Q), seamless integration of this framework with HPC resource managers (Slurm), and progressively tighter hardware-level integration between classical and quantum resources to minimize latency. Co-locating the QPUs with the classical resources is a first step towards minizing latency, and we anticipate that further integrating the classical resources with the QPUs control systems will allow us to achieve a sub-millisecond round-trip latency. This tight integration will be crucial for quantum error correction and for useful hybrid classical-quantum algorithms with fast feedback loops.

Another characteristic of hybrid classical-quantum architectures is that the heterogeneity of quantum hardware will continue to grow. Currently, CUDA-Q supports quantum hardware based on trapped ions, neutral atoms, superconducting qubits, photonics, and other types through its integration with aggregators such as Amazon Braket. CUDA-Q will continue growing and supporting more quantum hardware modalities and increasing its simulation capabilities.

Building on this first demonstration of such an environment, future work will focus on:
\begin{itemize}
\item Evaluating scheduling strategies to balance HPC performance objectives with quantum resource constraints;
\item Reducing latency for tightly coupled algorithms and exploring loose coupling approaches where suitable;
\item Exploring alternative CPU/GPU/QPU topologies, including mixed-modality quantum integration;
\item Extending CUDA-Q features for deeper heterogeneous integration;
\item Broadening the algorithmic scope to cover additional hybrid paradigms and real-world use cases;
\item Developing benchmark suites for hybrid systems that inform both performance tuning and scheduling policy.
\end{itemize}

By providing a functional, operationally realistic platform for hybrid computation, this work bridges the gap between conceptual integration models and deployable HPC–QC systems. We anticipate that this approach will accelerate both algorithmic innovation and systems-level co-design in pursuit of a practical quantum advantage.

\bibliography{references}

\begin{thebibliography}{10}

\bibitem{mcclean2016theory}
Jarrod~R McClean, Jonathan Romero, Ryan Babbush, and Al{\'a}n Aspuru-Guzik.
\newblock The theory of variational hybrid quantum-classical algorithms.
\newblock {\em New Journal of Physics}, 18(2):023023, 2016.

\bibitem{endo2021hybrid}
Suguru Endo, Zhenyu Cai, Simon~C Benjamin, and Xiao Yuan.
\newblock Hybrid quantum-classical algorithms and quantum error mitigation.
\newblock {\em Journal of the Physical Society of Japan}, 90(3):032001, 2021.

\bibitem{callison2022hybrid}
Adam Callison and Nicholas Chancellor.
\newblock Hybrid quantum-classical algorithms in the noisy intermediate-scale quantum era and beyond.
\newblock {\em Physical Review A}, 106(1):010101, 2022.

\bibitem{beck2024integrating}
Thomas Beck, Alessandro Baroni, Ryan Bennink, Gilles Buchs, Eduardo~Antonio Coello~Perez, Markus Eisenbach, Rafael Ferreira~da Silva, Muralikrishnan Gopalakrishnan~Meena, Kalyan Gottiparthi, Peter Groszkowski, Travis~S. Humble, Ryan Landfield, Ketan Maheshwari, Sarp Oral, Michael~A. Sandoval, Amir Shehata, In-Saeng Suh, and Christopher Zimmer.
\newblock Integrating quantum computing resources into scientific {HPC} ecosystems.
\newblock {\em arXiv preprint arXiv:2408.16159}, 2024.

\bibitem{elsharkawy2024integration}
Amr Elsharkawy, Xiaorang Guo, and Martin Schulz.
\newblock Integration of quantum accelerators into {HPC}: Toward a unified quantum platform.
\newblock {\em arXiv preprint arXiv:2407.18527}, 2024.

\bibitem{honda2025advantages}
Daigo Honda, Yuta Nishiyama, Junya Ishikawa, Kenichi Matsuzaki, Satoshi Miyata, Tadahiro Chujo, Yasuhisa Yamamoto, Masahiko Kiminami, Taro Kato, Jun Towada, Naoki Yoshioka, Naoto Aoki, and Nobuyasu Ito.
\newblock Advantages of co-locating quantum-{HPC} platforms: A survey for near-future industrial applications.
\newblock {\em arXiv preprint arXiv:2508.04171}, 2025.

\bibitem{shehata2025software}
Amir Shehata, Peter Groszkowski, Thomas Naughton, Murali Gopalakrishnan~Meena, Elaine Wong, Daniel Claudino, Rafael Ferreira~da Silva, and Thomas Beck.
\newblock Building a software stack for quantum-{HPC} integration.
\newblock {\em arXiv preprint arXiv:2503.01787}, 2025.

\bibitem{mantha2024pilot}
Pradeep Mantha, Florian~J. Kiwit, Nishant Saurabh, Shantenu Jha, and Andre Luckow.
\newblock Pilot-quantum: A quantum-{HPC} middleware for resource, workload and task management.
\newblock {\em arXiv preprint arXiv:2412.18519}, 2024.

\bibitem{zappin2024quantum}
Jake Zappin, Trevor Stalnaker, Oscar Chaparro, and Denys Poshyvanyk.
\newblock When quantum meets classical: Characterizing hybrid quantum-classical issues discussed in developer forums.
\newblock {\em arXiv preprint arXiv:2411.16884}, 2024.

\bibitem{bensoussan2025taxonomy}
Avner Bensoussan, Gunel Jahangirova, and Mohammad~Reza Mousavi.
\newblock A taxonomy of real faults in hybrid quantum-classical architectures.
\newblock {\em arXiv preprint arXiv:2502.08739}, 2025.

\bibitem{mcclean2016variational}
Ryan~Babbush Jarrod R~McClean, Jonathan~Romero and Alán Aspuru-Guzik.
\newblock The theory of variational hybrid quantum-classical algorithms.
\newblock In {\em New Journal of Physics}, page Volume 18, 2016.

\bibitem{gujju2024quantum}
Yaswitha Gujju, Atsushi Matsuo, and Rudy Raymond.
\newblock Quantum machine learning on near-term quantum devices: Current state of supervised and unsupervised techniques for real-world applications.
\newblock {\em Physical Review Applied}, 21(6):067001, 2024.

\bibitem{yoo2003slurm}
Andy~B Yoo, Morris~A Jette, and Mark Grondona.
\newblock Slurm: Simple linux utility for resource management.
\newblock In {\em Workshop on job scheduling strategies for parallel processing}, pages 44--60. Springer, 2003.

\bibitem{Kurowski2004}
Krzysztof Kurowski, Jarek Nabrzyski, Ariel Oleksiak, and Jan W{\k{e}}glarz.
\newblock {\em Multicriteria Aspects of Grid Resource Management}, pages 271--293.
\newblock Springer US, Boston, MA, 2004.

\bibitem{britt2017high}
Keith~A Britt and Travis~S Humble.
\newblock High-performance computing with quantum processing units.
\newblock {\em ACM Journal on Emerging Technologies in Computing Systems (JETC)}, 13(3):1--13, 2017.

\bibitem{novak2024boundaries}
Samo Nov{\'a}k, David~D Roberts, Alexander Makarovskiy, Ra{\'u}l Garc{\'\i}a-Patr{\'o}n, and William~R Clements.
\newblock Boundaries for quantum advantage with single photons and loop-based time-bin interferometers.
\newblock {\em arXiv preprint arXiv:2411.16873}, 2024.

\bibitem{aaronson2011computational}
Scott Aaronson and Alex Arkhipov.
\newblock The computational complexity of linear optics.
\newblock In {\em Proceedings of the forty-third annual ACM symposium on Theory of computing}, pages 333--342, 2011.

\bibitem{bacarreza2025quantum}
Omar Bacarreza, Thorin Farnsworth, Alexander Makarovskiy, Hugo Wallner, Tessa Hicks, Santiago Sempere-Llagostera, John Price, Robert~J.A Francis-Jones, and William Clements.
\newblock Quantum latent distributions in deep generative models.
\newblock {\em manuscript in preparation}, 2025.

\bibitem{cimini2025large}
Valeria Cimini, Mandar~M Sohoni, Federico Presutti, Benjamin~K Malia, Shi-Yuan Ma, Ryotatsu Yanagimoto, Tianyu Wang, Tatsuhiro Onodera, Logan~G Wright, and Peter~L McMahon.
\newblock Large-scale quantum reservoir computing using a {Gaussian} boson sampler.
\newblock {\em arXiv preprint arXiv:2505.13695}, 2025.

\bibitem{pcss2025hybriddemo}
Mateusz Slysz, Łukasz Grodzki, Piotr Rydlichowski, Dawid Siera, Krzysztof Kurowski, Grzegorz Waligóra, and Jan Węglarz.
\newblock Solving combinatorial optimization and machine learning problems on hybrid near-term quantum photonic computers.
\newblock In {\em Future Generation Computer Systems}, page Volume 174, 2025.

\bibitem{gan2022fock}
Beng~Yee Gan, Daniel Leykam, and Dimitris~G Angelakis.
\newblock Fock state-enhanced expressivity of quantum machine learning models.
\newblock {\em EPJ Quantum Technology}, 9(1):16, 2022.

\bibitem{cyber_benchmark}
Nour Moustafa and Jill Slay.
\newblock {UNSW-NB15}: a comprehensive data set for network intrusion detection systems.
\newblock In {\em 2015 Military Communications and Information Systems Conference (MilCIS)}, pages 1--6, 2015.

\bibitem{kornjavca2024large}
Milan Kornja{\v{c}}a, Hong-Ye Hu, Chen Zhao, Jonathan Wurtz, Phillip Weinberg, Majd Hamdan, Andrii Zhdanov, Sergio~H Cantu, Hengyun Zhou, Rodrigo~Araiza Bravo, et~al.
\newblock Large-scale quantum reservoir learning with an analog quantum computer.
\newblock {\em arXiv preprint arXiv:2407.02553}, 2024.

\bibitem{gong2025enhanced}
Si-Qiu Gong, Ming-Cheng Chen, Hua-Liang Liu, Hao Su, Yi-Chao Gu, Hao-Yang Tang, Meng-Hao Jia, Yu-Hao Deng, Qian Wei, Hui Wang, et~al.
\newblock Enhanced image recognition using {Gaussian} boson sampling.
\newblock {\em arXiv preprint arXiv:2506.19707}, 2025.

\bibitem{yin2024experimental}
Zhenghao Yin, Iris Agresti, Giovanni de~Felice, Douglas Brown, Alexis Toumi, Ciro Pentangelo, Simone Piacentini, Andrea Crespi, Francesco Ceccarelli, Roberto Osellame, et~al.
\newblock Experimental quantum-enhanced kernels on a photonic processor.
\newblock {\em arXiv preprint arXiv:2407.20364}, 2024.

\bibitem{kailasanathan2025quantum}
Rajiv Kailasanathan, William Clements, Mohammad~Reza Boskabadi, Shawn~M. Gibford, Emmanouil Papadakis, Christopher~J. Savoie, and Seyed~Soheil Mansouri.
\newblock Quantum enhanced ensemble {GANs} for anomaly detection in continuous biomanufacturing.
\newblock {\em manuscript in preparation}, 2025.

\bibitem{facelli2024exact}
Giorgio Facelli, David~D Roberts, Hugo Wallner, Alexander Makarovskiy, Zo{\"e} Holmes, and William~R Clements.
\newblock Exact gradients for linear optics with single photons.
\newblock {\em arXiv preprint arXiv:2409.16369}, 2024.

\bibitem{hoch2025variational}
Francesco Hoch, Giovanni Rodari, Taira Giordani, Paul Perret, Nicol{\`o} Spagnolo, Gonzalo Carvacho, Ciro Pentangelo, Simone Piacentini, Andrea Crespi, Francesco Ceccarelli, et~al.
\newblock Variational approach to photonic quantum circuits via the parameter shift rule.
\newblock {\em Physical Review Research}, 7(2):023227, 2025.

\bibitem{chen2025distributed}
Kuan-Cheng Chen, Chen-Yu Liu, Yu~Shang, Felix Burt, and Kin~K Leung.
\newblock Distributed quantum neural networks on distributed photonic quantum computing.
\newblock {\em arXiv preprint arXiv:2505.08474}, 2025.

\end{thebibliography}
\bibliographystyle{unsrt}

\end{document}